\documentclass[twocolumn,showpacs,preprintnumbers,amsmath,amssymb]{revtex4}

\usepackage{graphicx}
\usepackage{dcolumn}
\usepackage{bm}

\begin{document}

\preprint{}

\title{Entangled photons from on-chip slow light}

\author{Hiroki Takesue$^{1}$} \email{takesue.hiroki@lab.ntt.co.jp}
\author{Nobuyuki Matsuda$^{1,2}$} 
\author{Eiichi Kuramochi$^{1,2}$} 
\author{Masaya Notomi$^{1,2}$} 
\affiliation{%
$^1$NTT Basic Research Laboratories, NTT Corporation, 3-1 Morinosato Wakamiya, Atsugi, Kanagawa, 243-0198, Japan\\
$^2$Nanophotonics Center, NTT Corporation, 3-1 Morinosato Wakamiya, Atsugi, Kanagawa, 243-0198, Japan
}%

\date{\today}

\begin{abstract}
We report the first entanglement generation experiment using an on-chip slow light device. With highly efficient spontaneous four-wave mixing enhanced by the slow light effect in a coupled resonator optical waveguide based on a silicon photonic crystal, we generated 1.5-$\mu$m-band high-dimensional time-bin entangled photon pairs. We undertook two-photon interference experiments and observed the coincidence fringes with visibilities $>74$\%. The present result enables us to realize an on-chip entanglement source with a very small footprint, which is an essential function for quantum information processing based on integrated quantum photonics.
\end{abstract}

\maketitle

Slow light waveguides are now being intensively studied as a way of realizing various functionalities such as optical switches and buffers with very small footprints on a chip \cite{krauss,baba}. Slow light waveguides are also attracting attention as a way of obtaining a huge nonlinear optical effect. For example, the third order nonlinear coefficient $\gamma$ is proportional to the square of the group index $n_{g}$ \cite{sol}, which means that we can significantly enhance the four-wave mixing efficiency by using a slow light waveguide \cite{monat}. Thus, a slow light waveguide is a promising candidate as a nonlinear optical device for all-optical signal processing. 
The slow-light enhanced nonlinearity is also useful for integrated quantum photonics, in which photonic quantum information processing functions are integrated on an optical waveguide \cite{plc}. With the aim of realizing integrated photon sources on a chip, several correlated photon pair generation experiments based on slow-light enhanced spontaneous four-wave mixing (SFWM) have already been reported \cite{xiong,davanco,correlated}. Xiong et al. demonstrated the first correlated pair generation enhanced by a slow light waveguide, using a dispersion-engineered silicon photonic crystal waveguide \cite{xiong}. This experiment was followed by one that used a coupled resonator optical waveguide (CROW) \cite{yariv} based on silicon ring cavities as a slow-light device \cite{davanco}. Recently, our group reported a correlated pair generation experiment using a CROW based on silicon photonic crystal cavities \cite{correlated}. These experiments achieved a significant enhancement of pair generation efficiency compared with conventional photon pair sources based on SFWM. However, none of the experiments have achieved the generation of an entangled state, or in other words, confirmed the coherence of two-photon states. 

In this paper, we report the first entanglement generation experiment to use a slow-light waveguide. We utilized slow-light enhanced SFWM in a CROW consisting of 200 silicon photonic crystal nanocavities \cite{notomi} pumped by a coherent pulse train to generate a high-dimensional time-bin entangled state \cite{sequential}. We undertook two-photon interference measurements using a Franson interferometer \cite{franson} and observed the coincidence fringes with visibilities exceeding the limit of the violation of Bell's inequality. 

\section*{Results} 

\subsection*{CROW based on silicon photonic crystal}

\begin{figure*}[bht]

\centerline{\includegraphics[width=\linewidth]{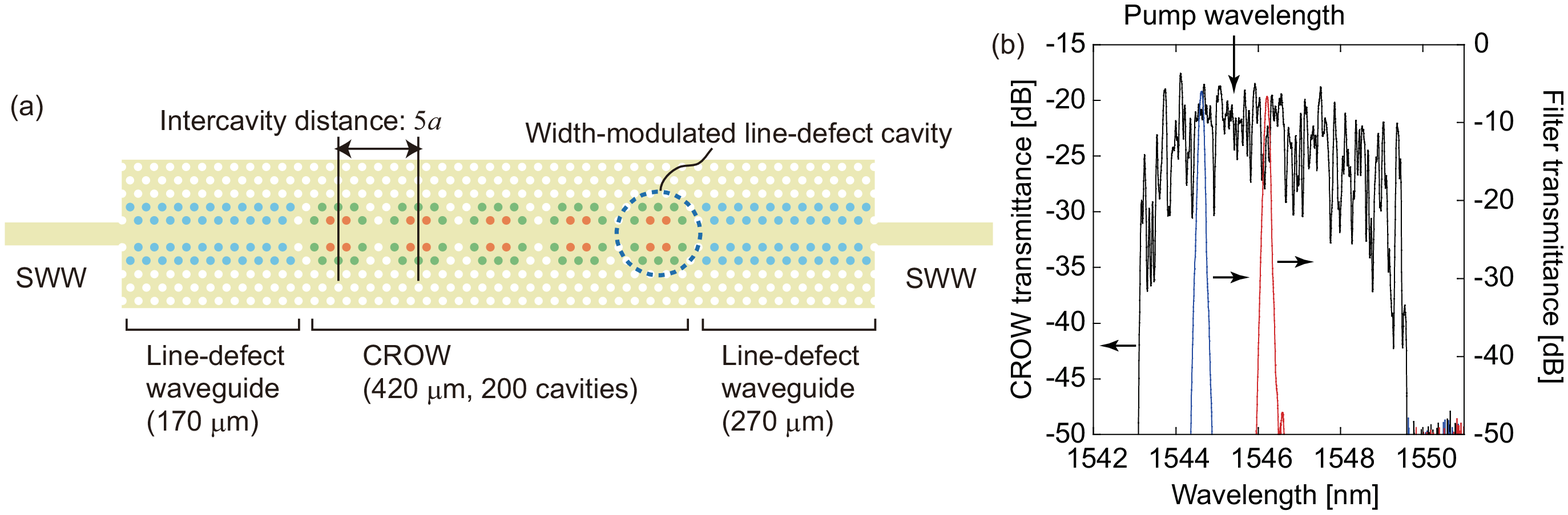}}

\caption{(a) CROW used in our experiments. (b) Fiber-to-fiber transmission spectra of CROW (black line) at 32.8$^\circ$C and filter transmittances for the signal (blue) and idler (red) channels. The position of the pump wavelength is indicated by an arrow. The resolution bandwidth is 0.02 nm. } \label{1}

\end{figure*}

A CROW is a one-dimensional array of identical optical cavities, where adjacent cavities are coupled to each other to form an extended mode along the waveguide. A CROW exhibits a transmission bandwidth that is far larger than those of the individual cavities, while the group velocity is significantly reduced inside the band. Figure \ref{1} (a) shows a schematic of our CROW fabricated on a silicon chip. The CROW is based on a width-modulated line defect cavity in a silicon photonic crystal (PhC) with a two-dimensional triangular-lattice of air holes \cite{kuramochi}. The $Q$ of each cavity is $\sim 10^6$, and the intercavity distance is $5a$, where $a$ is the lattice constant of our PhC waveguide (420 nm). The number of cavities is 200, which means that the total length of the CROW is 420 $\mu$m. PhC line-defect waveguides are equipped at the both ends of the CROW. The PhC section is integrated with silicon wire waveguides (SWW) from which we can optically access the CROW. Lensed fibers are used to couple light to the silicon chip with a coupling efficiency of $\sim -9$ dB per point. Figure \ref{1} (b) shows the transmission spectra of the CROW at a temperature of 32.8$^\circ$C. The formation of a broad transmission band was clearly observed from 1543 to 1550 nm. 

\subsection*{Experiments}

\begin{figure}[bht]

\centerline{\includegraphics[width=\linewidth]{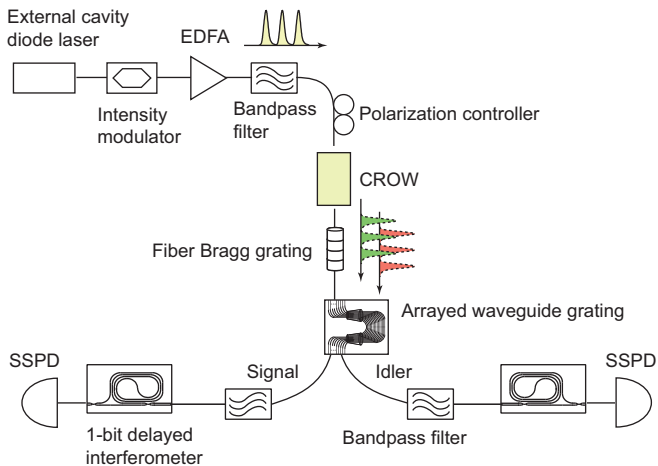}}

\caption{Experimental setup} \label{2}
\vspace{5mm}
\end{figure}

Figure \ref{2} shows the experimental setup. A 1545.4-nm continuous wave light from an external cavity laser diode was modulated into a 60-ps, 1-GHz clock pulse train using a lithium niobate intensity modulator. The pulse train was then amplified by an erbium-doped fiber amplifier (EDFA), and input into optical bandpass filters to eliminate the amplified spontaneous emission noise from the EDFA. The pulse train was launched into the silicon chip that contained the CROW through a lensed fiber. The chip temperature was set at 32.8$^\circ$C. At this temperature, the group index of the CROW for the pump wavelength was around 40 \cite{fwm}. Through the slow-light enhanced SFWM in the CROW, high-dimensional time-bin entangled photon pairs were generated, whose quantum state is approximately given by \cite{sequential}\begin{equation}
|\Psi\rangle = \sum_{k=1}^N |k\rangle_s |k\rangle_i. \label{ent}
\end{equation}
Here $N$ is the number of pump pulses where coherence is preserved, and $|k\rangle_x$ denotes a quantum state where there is a photon at the $k$th time slot in the mode $x$ ($=s$: signal, $i$: idler). The photons from the chip were collected by another lensed fiber, passed through fiber Bragg gratings to suppress the pump photons, and input into an arrayed waveguide grating to separate the signal and idler channels. The wavelengths of the signal and idler channels were 1544.6 and 1546.2 nm, respectively, and both had a 0.1 nm (12.5 GHz) bandwidth. 
The transmission spectra of the signal and idler channels with respect to the CROW band are plotted in Fig. \ref{1} (b). The total channel losses were 6.0 and 6.7 [dB] for the signal and idler, respectively. 
The signal and idler photons were passed through optical bandpass filters to suppress the pump photons further, and launched into 1-bit delayed Mach-Zehnder interferometers made using silica-on-silicon waveguides \cite{timebin}. The phase differences between the two arms of the interferometers were tuned by changing the waveguide temperature: a $\pi$ phase shift was attained with $\sim$ 0.4 $^\circ$C temperature change. 
With these interferometers, a state $|k\rangle_x$ is converted to $(|k\rangle_x + e^{i \phi_x} |k+1\rangle_x)/2$, where $\phi_x$ is the phase difference induced in the interferometer for channel $x$. Then, Eq. (\ref{ent}) is converted to 
\begin{equation}
|\Psi\rangle \to |1\rangle_s |1\rangle_i + \sum_{k=2}^{N} (1+e^{i(\phi_s + \phi_i)}) |k\rangle_s |k\rangle_i + |N+1\rangle_s |N+1\rangle_i,
\end{equation}
where only the terms that contributed to the coincidence detection are shown and the normalizing term is discarded for simplicity. 
By ignoring the first and the last terms on the right hand side of the above equation, we obtain the coincidence probability $P$ as  
\begin{equation}
P \propto 1 + V \cos (\phi_s + \phi_i), \label{coin}
\end{equation}
where $V$ is the two-photon interference visibility.
Then, the photons from the interferometers were received by superconducting single photon detectors (SSPD), and the coincidences were counted using a time interval analyzer. The detection efficiency of both SSPDs was set at 20\%, and at this condition, the dark count rates were 50 and 10 Hz for the signal and idler channels, respectively.

\begin{figure*}[htb]

\centerline{\includegraphics[width=\linewidth]{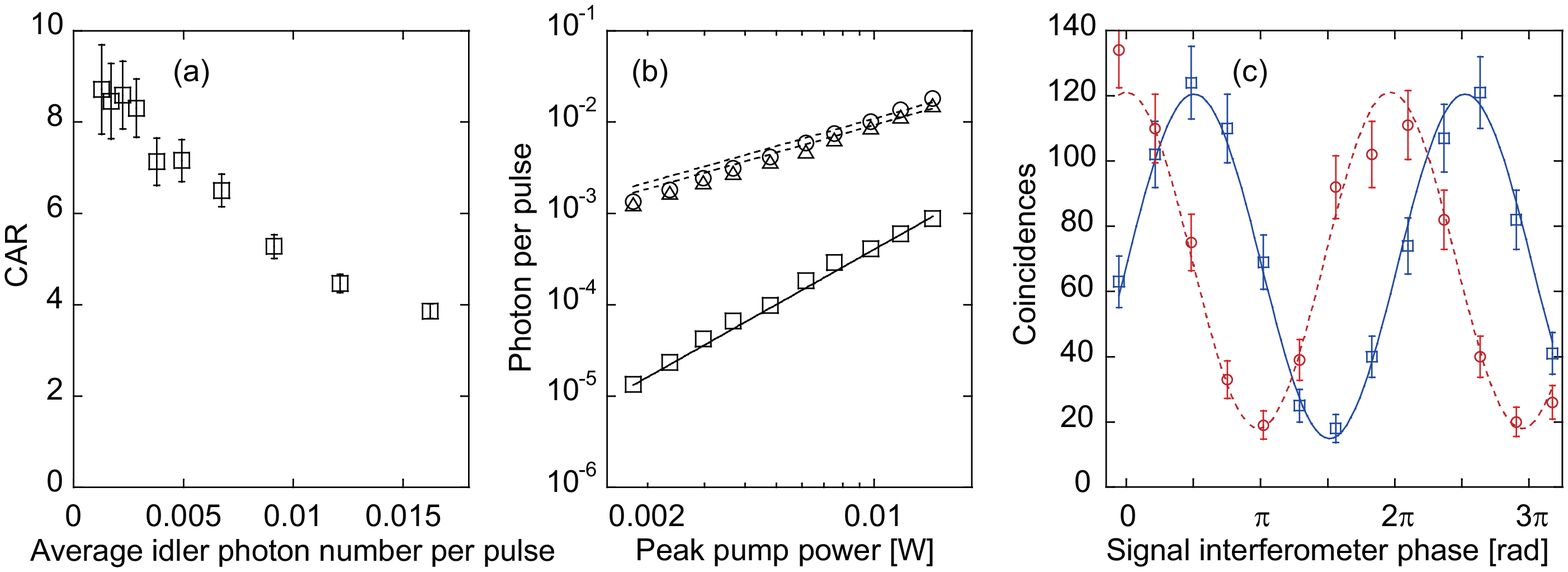}}

\caption{(a) CAR as a function of the average idler photon number per pulse. (b) Average numbers of correlated (squares) and noise photons for the signal (circles) and idler (triangles) per pulse as a function of peak pump power. The correlated and noise photon data are fitted with square (solid line) and linear (dotted lines) functions, respectively. (c) Two-photon-interference fringes. Squares: idler interferometer phase at 0, circles: $\pi/2$. Coincidences were counted for 10 million start pulses. } \label{3}
\vspace{5mm}
\end{figure*}

To investigate the degree of temporal correlation, we first removed the waveguide interferometers from the setup and undertook a coincidence-to-accidentals ratio (CAR) measurement. The result is shown in Fig. \ref{3} (a). We obtained the maximum CAR of $8.7 \pm 1.0$ when we set the average photon number per pulse at 0.001. The improvement in the CAR compared with our previous experiment \cite{correlated} can be mainly attributed to the fact that we halved the photon pair bandwidth and reduced the pump pulse width (details are provided in Method). 
Using the method reported in \cite{cool}, we can estimate the portions of correlated pairs and the noise photons from the CAR and count rate measurement results. The estimated correlated and noise photons per pulse as a function of peak coupled pump power are shown in Fig. \ref{3} (b). The average number of correlated photon pairs per pulse ($\mu_c$) is proportional to the square of the pump power, which is a clear signature of the SFWM process. 
With this result, we estimated that the effective nonlinear coefficient of our CROW was $\sim 5900$ [1/W/m], which is comparable to values reported for the previous experiments \cite{xiong,davanco,correlated} and larger than that of a conventional SWW by an order of magnitude \cite{silicon}. 
On the other hand, the noise photons were dominant in our experiment, which severely limited the maximum CAR. The average numbers of noise photons for the signal ($\mu_{sn}$) and the idler ($\mu_{in}$) are almost proportional to the pump power. Note that similar noise photon generation at small detunings was also observed in an SWW \cite{clemmen}. 

We then inserted waveguide interferometers into the setup and undertook a two photon interference experiment. The average photon number per pulse was set at 0.004. Figure \ref{3} (c) shows the coincidences as a function of the signal interferometer phase, where the squares and circles correspond to the data obtained when we set the idler interferometer phase at 0 and $\pi/2$, respectively. We observed clear sinusoidal modulations of coincidences for two nonorthogonal measurement bases for the idler photons, and the visibilities of the fitted curves were $78.0 \pm 3.9$\% (idler interferometer phase 0) and $74.1 \pm 4.8$\% ($\pi/2$). Thus, we confirmed the generation of a time-bin entangled state that can violate the Bell's inequality. 

\subsection*{Discussion}

As stated above, there were a significant number of noise photons, which limited the CAR and visibilities in the two photon interference measurements. Although the origin of the noise photons is not yet clear, the generation efficiency of the similar noise photons observed in an SWW exhibited temperature dependence, suggesting that those photons were generated by the scattering of the pump photons on thermal bath of excitations \cite{clemmen}. Therefore, we may be able to reduce the noise by cooling the device, which constitutes important future work. Also, as discussed in Method, we can improve the visibility by using narrower filters, at the expense of source brightness. 

The source presented here, whose device length is only 420 $\mu$m, is the smallest entanglement source yet reported. We can reduce the device footprint even further by optimizing the CROW parameters. For example, we can decrease the cavity coupling by employing a larger intercavity distance, which results in an increased $n_g$ \cite{yariv}.

In summary, we have reported the first entanglement generation using SFWM in an on-chip slow light device. Thanks to SFWM enhanced by the slow-light effect in a silicon PhC CROW, we realized an on-chip time-bin entangled photon pair source with a device length as short as 420 $\mu$m. The two-photon-interference experimental results showed that we could generate an entanglement that can violate Bell's inequality. 

\vspace{5mm}


\section*{Method}

\noindent 
{\bf Dependence of CAR on filter and pump pulse width.} 
In many previous SFWM experiments, noise photons caused by the spontaneous Raman scattering (SpRS) process degraded the quantum correlation. 
Inoue and Shimizu studied the effect of SpRS noise photons on photon pair generation based on SFWM with a pulsed pump \cite{kyo}. 
They pointed out that, at a fixed average photon number per pulse, the portion of SFWM photons increases as we reduce the filter bandwidth or the pump pulse width. Here we show the formula that relates CAR, filter bandwidth and pump pulse width. 

In the presence of noise photons whose number increases linearly with the pump power, the average photon number per pulse $\mu$ is expressed as follows. 

\begin{equation}
\mu = \mu_c (p) + \mu_n (p) \label{mu}
\end{equation}
Here, $p$ is the peak pump power and $\mu_c (p)$ and $\mu_n (p)$ are the average correlated and noise photon per pulse, which are given by
\begin{eqnarray}
\mu_c (p) &=& a p^2 \Delta f \Delta t, \\
\mu_n (p) &=& b p \Delta f \Delta t,
\end{eqnarray}
where $a$ and $b$ denote coefficients representing the efficiencies of SFWM and the noise-photon generation process, respectively. $\Delta f$ and $\Delta t$ are the filter bandwidth and the pump pulse width, respectively. In these equations, we assumed the numbers of noise photons to be the same for the signal and idler channels. Equation (\ref{mu}) implies that the peak pump power $p$ can be expressed with the following equation at a fixed $\mu$. 
\begin{equation}
p = -\frac{b}{2a} + \sqrt{\frac{1}{4} \left(\frac{b}{a}\right)^2+\frac{\mu}{a \Delta f \Delta t}} \label{p}
\end{equation}
On the other hand, assuming Poisson-distributed photon-pairs, CAR can be related to $\mu_c (p)$ and $\mu_n (p)$ with the following equation \cite{cool}. 
\begin{equation}
CAR= \frac{\mu_c (p) \alpha^2}{\{(\mu_c (p) + \mu_n (p)) \alpha + d\}^2} + 1 \label{car}
\end{equation}
Here, $\alpha$ and $d$ denote the coupling efficiency (including detection efficiency) of photons and the dark count probability of the detectors, respectively, assuming that these values are the same for both signal and idler channels. 
With Eqs. (\ref{p}) and (\ref{car}), we obtain the following equation. 
\begin{equation}
CAR = \left(\frac{\mu \alpha}{\mu \alpha + d}\right)^2 \frac{4 a}{\Delta f \Delta t\left(b+\sqrt{b^2 + \frac{4 a \mu}{\Delta f \Delta t}} \right)^2}+1 \label{car2}
\end{equation}
It is clear that we can observe a better CAR if we reduce $\Delta f \Delta t$. 

With the experimental data shown in Fig. \ref{3} (b), we obtain $a=5.78$ and $b=1.03$ at $\mu=0.004$. 
In our previous experiment reported in \cite{correlated}, $\Delta f$, $\Delta t$ and $d$ were 25 [GHz], 100 [ps] and $7 \times 10^{-6}$, respectively.  
If we use these values together with $a$ and $b$ shown above to calculate CAR using Eq. (\ref{car2}), we obtain 2.2, which is close to the value presented in \cite{correlated} ($\sim 2$). On the other hand, the calculated CAR is 3.1 if we use the SSPDs with the filter and pump pulse width shown above. This consideration implies that the CAR enhancement in the present experiment can be mainly attributed to the reduction of the filter and pump pulse widths.

\section*{Acknowledgements}

This work was supported by the Japan Society for the Promotion of Science with a Grant-in-Aid for Scientific Research (No. 22360034).


\section*{Author contributions} 

H.T. and N.M. performed the experiments and data analysis. 
M.N. and E.K. designed and fabricated the CROW device. 
H.T. managed the project. 
All authors discussed the results and wrote the manuscript.

\section*{Additional information}

{\bf Competing financial interests:} The authors declare that there are no competing financial interests.














\end{document}